\documentclass[english,aps,prb,showpacs,twocolumn]{revtex4}
\usepackage{graphicx}
\usepackage{babel}
 
\begin{document}

\title{Electronic continuum states and far infrared absorption of InAs/GaAs quantum dots}

\author{D.~P.~Nguyen}

\author{N.~Regnault}

\author{R.~Ferreira}

\author{G.~Bastard}

\affiliation{Laboratoire Pierre Aigrain - Ecole Normale Sup\'erieure,
24 rue Lhomond, F-75005 Paris, France}

\begin{abstract}

The electronic continuum states of InAs/GaAs semiconductor quantum dots embedded in a GaAs/AlAs superlattice are theoretically investigated and the far/mid infrared absorption spectra are calculated for a variety of structures and polarizations. The effect of a strong magnetic field applied parallel to the growth direction is also investigated. We predict that the flatness of the InAs/GaAs dots leads to a mid-infrared absorption which is almost insensitive to the magnetic field, in spite of the reorganization of the continuum into series of quasi-Landau states. We also predict that it is possible to design InAs/GaAs photoconductors which display very strong in-plane absorption.
\end{abstract}


\maketitle

\section{Introduction}

It is by now well established that the lowest lying eigenstates of the InAs/GaAs quantum dots (QDs) are bound states if the dots are not too small. At higher energies one finds continuum states which are either bulk-like (i.e. display extended wave functions in the three directions of space) or are bound along the growth axis, the wetting layer states \cite{Bimberg99}. The continuum states are of paramount importance to the control of useful electronic properties associated with the bound states (e.g. lasing action or far infrared photoresponse). This is because their density of states are considerably larger than that of the bound states of the dots~since the QDs are most of the time very diluted. Despite their importance, there have been relatively few calculations dealing with the QD's continuum states \cite{Bassani96, Lelong00, Schrey04, Vasanelli01}. This is probably because the non trivial shape of the dots renders the continuum states hardly analytically solvable. Spherical dot virtual bound states were discussed by Buczko and Bassani \cite{Bassani96} while Lelong et al \cite{Lelong00} related the photoconductivity properties of lens-shaped QDs to the existence of virtual bound states. 

In the present paper we examine the structure of the electronic continuum states of the InAs/GaAs QDs. We shall show that the flatness of the QDs reorganizes very effectively the continuum states. In fact, it is the dominant feature that influences the far infrared response \cite{Rebohle02, Fricke96, Maimon98, Lee99, Sauvage97, Hameau99, Isaia02}. This response is associated with the photon absorption from the ground state to the continuum states. We shall show in particular that a strong magnetic field applied parallel to the growth axis leads to the formation of quasi-Landau states, as expected. However, this quantization is almost invisible in the far infrared response with light polarization along the growth direction. Finally, we shall show that one can design InAs/GaAs photodetectors which display very strong dependence of their photoabsorption upon the in-plane polarization of the light.

\section{Model}

The InAs/GaAs self-organized dots are known to be flat objects ($h/R \sim$~2-3~nm/10~nm where $h$ is the dot height and $R$ the base radius). It is also known that they are roughly circular. The energy distance between the $S$-like symmetry envelope ground state in the conduction band and the center of gravity of the $P_x$ and $P_y$ first excited states is about 50~meV for $h$ about 2~nm and $R$ about 10~nm \cite{Schrey04, Fricke96, Hameau99, Isaia02}. The splitting between $P_x$ and $P_y$ (about 5~meV) results from potential energy terms which do not display cylindrical symmetry, for instance ellipticity of the QD basis \cite{Hameau99} or/and piezo-electric fields \cite{Stier99, Stier01}. The electronic structure of QDs by means of the multi-band envelope function method has been calculated by Stier et al \cite{Stier99, Stier01} while accurate atomistic approaches (pseudo-potential, tight-binding) were undertaken by Williamson and Zunger \cite{Zunger00} (see also Zunger \cite{Zunger02} for a review) and by Lee et al \cite{Lee01}. These numerical methods (of either type) mostly aimed at calculating the bound states of the dots. Here, we shall use the simpler (one band) envelope function description to handle the more complicated extended states of the dots and to predict the QD's far infrared response. We note that this one band envelope function method has allowed a quantitative description of numerous anti-crossings observed in the fan charts of the bound-to-bound magneto-optical transitions of InAs/GaAs QDs and associated with polarons \cite{Hameau99, Isaia02}. 
The effective Hamiltonian we investigate is :
\begin{eqnarray}
H(B) &=& H_0(B) + \delta V(\vec{r}) \\
     &=& \frac{p^2}{2 m^*} + V(\rho, z) + \frac{1}{2} \omega _c L_z + \frac{1}{8} m^* \omega_c ^2 \rho ^2 + \delta V(\vec{r}) \nonumber
\label{Hamiltonian}
\end{eqnarray}

where $m^*$ is the carrier effective mass (taken as constant), $V(\rho, z)$ the isotropic part of the potential energy, $\omega _c = eB/m^*$ the cyclotron frequency and $\delta V$ the potential energy part which does not display the cylindrical symmetry. $B$ is the magnitude of the magnetic field applied parallel to the $z$ direction. The eigenstates of $H_0$ can be chosen as eigenfunctions of $L_z$ with the eigenvalue $m$: they will be termed $S, P_+, P_-, D_+, D_-,~\dots$. A single quantum dot plane faintly absorbs light. Hence, to increase absorption, one very often uses vertically stacked dot planes. When a periodic stacking with a short period $d$ (say $d = 11$~nm) is used \cite{Rebohle02}, a vertical alignment of the QDs is obtained due to their strain distribution \cite{Hofer02}. Hence, one effectively structures the QD's continuum. This corresponds to looking for Bloch-like solutions of $H(B)$:
\begin{equation}
\psi_{k_z}(x, y, z + d) = e^{\imath k_z d} \cdot \psi_{k_z}(x, y, z)
\label{Bloch}
\end{equation}

The eigenenergies in turn become periodic functions of $k_z$. The first Brillouin zone will be the segment [$-\pi / d$,~$+\pi /d$]. The beneficial action on the photoresponse of QDs of such vertically periodic stackings with the insertion of AlAs barriers in the unit cells has been recently investigated \cite{Rebohle02}. One possible use of QDs arrays is the detection of infrared light. Under such circumstances the dots are modulation-doped and the doping concentration is adjusted in such a way that each QD contains one electron. The far infrared light is absorbed by the QDs because of bound-to-bound transitions (e.g. $S \to P\pm$) or bound-to-continuum transitions. Two possible polarizations may arise : either the electric vector $\vec{\mathcal{E}}$ of the electromagnetic wave is parallel to the $z$ axis or $\vec{\mathcal{E}}$ lays in the layer plane. In the first situation, only the continuum states with an $S$ symmetry will be involved in the absorption processes. In the second situation, only the continuum states with a $P_\pm$ symmetry will be involved in the response to the electromagnetic perturbation (cylindrical symmetry) or, more generally, the linear combinations of $P_+$ and $P_-$ which diagonalizes $H$. In the following, we shall present the absorption coefficient $\mathcal{P}$ versus the photon energy $\hbar \omega$ for different polarizations:
\begin{equation}
\mathcal{P}(\hbar \omega)\propto\sum_\nu \frac{2m^* \omega^2}{\hbar \omega} |\langle S|\vec{\mathcal{E}}\cdot\vec{r}|\nu\rangle|^2\delta(\epsilon_\nu-\epsilon_S- \hbar\omega)
\end{equation}

where $\nu$ labels the continuum states. In practice, we broadened the delta function by replacing it by a Lorentzian of 8~meV full width at half maximum. The QD size distribution is known to broaden the absorption lines : an average of $\mathcal{P}$ over the size distribution of the dot $(R, h)$ has to be performed to allow a complete comparison to experiments. But the broadening effect remains modest for intra-conduction band transitions (a few meV) and can easily be taken into account \cite{Isaia02}, if necessary.

A convenient model which takes advantage of the flat aspect of the QDs is the separable model \cite{Vasanelli01}. It will prove very useful to interpret the more accurate results obtained by the numerical diagonalization (see below). In the separable model, restricted to the cylindrically symmetrical case for simplicity, the wave function is written as :
\begin{eqnarray}
\psi_m(\vec{\rho}, z) &=& e^{\imath m\varphi} \cdot g_m(\rho, z) \\
g_m(\rho, z) &=& N y_m(z) f_m(\rho) 
\end{eqnarray}
				
where $N$ is a normalization coefficient, $f_m(\rho)$ a prescribed function of $\rho$ which depends on several variational parameters $(\lambda _1, \lambda _2, \dots)$ and $y_m(z)$ an unknown function which describes the carrier kinematics along the growth axis as resulting from an average potential $V_m(z)$:
\begin{equation}
V_m(z) = N^2 \int 2\pi \rho d\rho |f_m(\rho)|^2 V(\rho, z)
\end{equation}

$V_m(z) = V_m(z + d)$ depend on $\lambda _1, \lambda _2, \dots$. The one dimensional Schr\"odinger equation for the $z$ motion is solved numerically. Hence, the eigenvalues of the decoupled problem depends on $\lambda _1, \lambda _2, \dots$ The lowest bound represents the best possible separable solutions. To take an example, suppose we discuss the $S$ states : $m = 0$. We try a one parameter Gaussian ansatz for $f_0$:
\begin{equation}
f_0(\rho) = \exp \left(-\frac{\rho ^2}{2 \lambda ^2}\right)
\label{GaussianAnsatz}
\end{equation}								

For a single QD, the lowest eigenvalue (corresponding to $\lambda _{min}$) will correspond to the ground bound state with $S$ symmetry. However, the Schr\"odinger equation for the $z$ motion with $V_{m=0}(z, \lambda _{min})$ admits excited solutions which can be bound to the QD or belong to the continuum spectrum. In the latter case, the decoupling procedure amounts to producing a particular set of continuum states, those which are bound and nodeless in the radial direction but are extended along the growth axis. Strictly speaking, these extra solutions should be considered with caution because they are not the ground state solutions to the problem. However, if the problem were truly separable, the variational ansatz would have no consequence; the effective potential for the $z$ motion would be $m$ independent and all the solutions of the $z$ dependent problem would be acceptable. Therefore, in the case of flat objects such as the QDs, the excited states of the effective problem for the $z$ motion may be close to actual quantum states of the problem. As we shall see in the following they are indeed very useful guidelines. Note that in the case of periodic stacking, the excited separable solutions will be found to display a significant integrated probability of being outside the QD in the unit cell of the periodic problem despite the localized nature of their lateral motion (as in eq. \ref{GaussianAnsatz}). 

The Gaussian ansatz described above represents the $1S$-like states as in atomic physics convention. An orthogonal ensemble of decoupled variational ansatz of same ($2S$, $3S$, $\dots$) or different ($1P_\pm$, $2P_\pm$, $\dots$) symmetry as the ground $1S$-like one can be easily constructed and their eigenenergies are determined following the same minimization procedure as for the first one. For instance, we chose the following ansatz for the $2S$ states which have one node in the plane:
\begin{equation}
f_0(\rho) = (\rho^2 - \rho_0^2) \exp \left(-\frac{\rho ^2}{2 \lambda_{2S} ^2}\right)
\label{2SAnsatz}
\end{equation}
The $\rho_0$ value is related to the parameter $\lambda_{2S}$ by imposing the orthogonality to the $1S$ states. Hence, the independent parameter $\lambda_{2S}$ was found by minimizing the first calculated eigenenergy.

\begin{figure}[!htbp]
\begin{center}\includegraphics[width=\columnwidth, keepaspectratio]{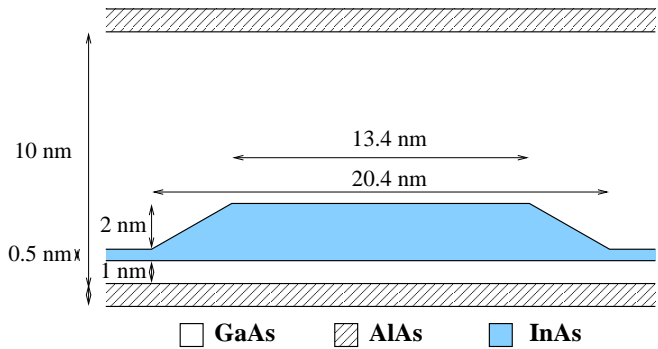}
\end{center}
\caption{Schematic representation of the supercell including the dot and its WL. Period $d = 11~\mathrm{nm}$.}
\label{dotmodel}
\end{figure}

In the following, we model the QDs as InAs truncated cones with basis radius $R$, height $h~=~2~\mathrm{nm}$, a basis angle $30^\circ$ (see fig. \ref{dotmodel}). The value of $R$ is taken equal to 10.2~nm if not mentioned differently. The cones float on a two-monolayer (0.5~nm) WL. The dot plane is placed at a distance of 1~nm from a 1~nm thick AlAs barrier. The overall period is $d = 11~\mathrm{nm}$. This sample was designed and investigated experimentally recently \cite{Rebohle02}.

The Bloch eigenstates and eigenvalues of a given $m$ (in case $\delta V = 0$) are found by projecting the Schr\"odinger equation on a large basis. At zero field, it is a Bessel basis for the radial motion \cite{Lelong00}. It corresponds to enclosing the QD in a large cylindrical box (100~nm radius) and requiring the wave function to vanish at the boundary. Plane wave functions with period $d$ are used to describe the vertical motion. At high magnetic fields ($B \ge 10$~T), the Bessel basis is replaced by a set of 2D harmonic oscillator functions (Landau levels). Altogether, we typically used~10000 basis functions for the Fourier-Bessel basis (21 plane wave functions and 500 Bessel functions) and 2800 basis functions for the Fourier-Landau basis (41 plane wave functions and 70 bidimensional harmonic oscillator functions). The Lanczos algorithm was used to extract the 30 lowest lying eigenstates. All calculations use a 0.4~eV \cite{Stier01} conduction band offset between GaAs and InAs and 1.08~eV \cite{Danan87} between AlAs and GaAs. The conduction band effective mass is taken equal to $0.07~m_0$ (viz. $\hbar \omega _c = 49.3~\mathrm{meV}$ if $B = 30~\mathrm{T}$) as determined by far infrared magneto-absorption experiments \cite{Hameau99}. The energy origin is taken at the bottom of the GaAs conduction band. We suppose that there is one electron per QD coming from doped layers. We also assume that the temperature is low enough to take for granted that all the electrons are in the ground states of the dots (the ground miniband of the stack is dispersionless because of the strong localization of the ground state of the individual QDs). Consequently, all the absorption processes are due to the excitation of these electrons.

\section{Results and discussions}

\subsection{$S$ states and excitation in $z$ polarization}
\label{Section1}

\begin{figure}[!htbp]
\begin{center}\includegraphics[width=\columnwidth, keepaspectratio]{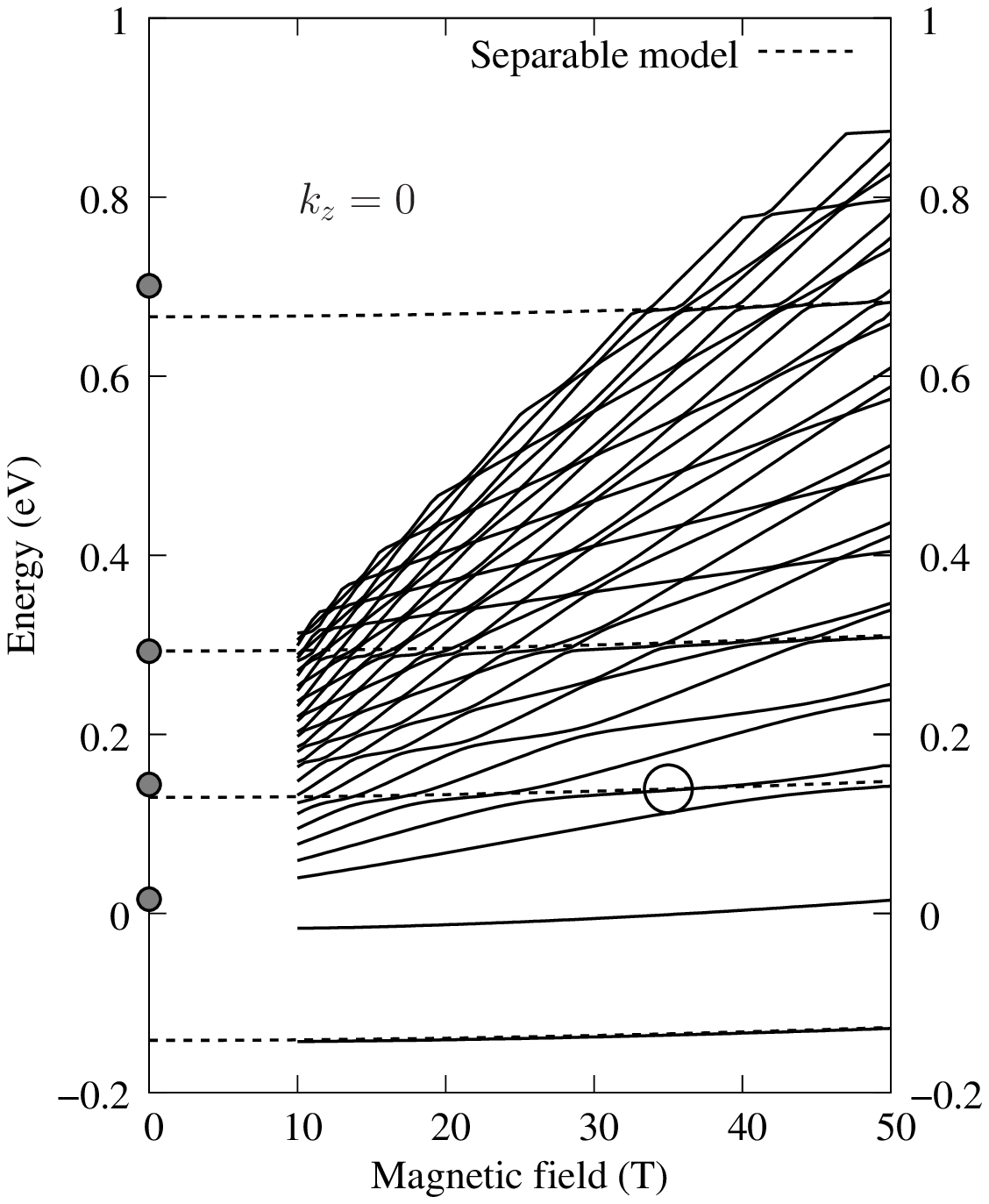}
\end{center}
\begin{center}\includegraphics[width=\columnwidth, keepaspectratio]{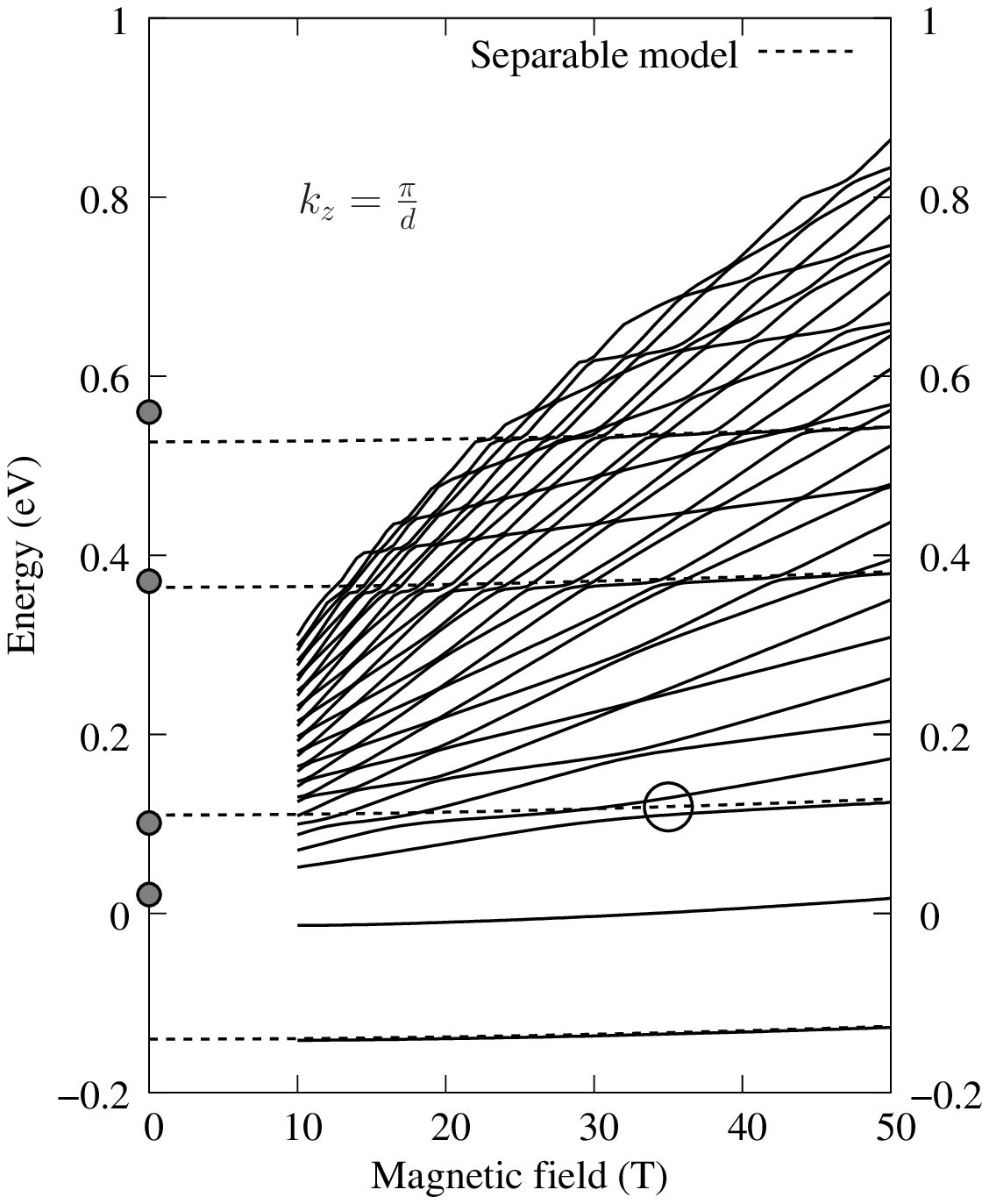}
\end{center}
\caption{Calculated energy levels of S symmetry states versus magnetic field. The dashed lines are the results of the separable model with a Gaussian radial function. Upper panel: at the center of the first Brillouin zone, lower panel: at an edge of the first Brillouin zone. The large dots at $B = 0$ are the extrapolations of the various fan.}
\label{S_states}
\end{figure}

Figure \ref{S_states} shows the calculated levels at $k_z = 0$ and $k_z = \pi / d$ for $S$ states at large field (full lines). We see two lowest lying isolated states followed by a series of levels organized in fan charts with extrapolations~at zero field: 16~meV, 144~meV, 293~meV, 701~meV for $k_z = 0$ and 21~meV, 101~meV, 371~meV, 560~meV for $k_z = \pi / d$ (large dots in fig. \ref{S_states}). Let us emphasize that these edges are roughly those of the GaAs/AlAs/InAs superlattice (i.e. with WL but without QDs). With our parameters such a superlattice (SL) would display $k_z=0~(\pi/d)$ edges at $E_1~=~19~\mathrm{meV}$ (31~meV), $E_2~=~151~\mathrm{meV}$ (110~meV), $E_3~=~293~\mathrm{meV}$ (374~meV) and $E_4~=~699~\mathrm{meV}$ (559~meV). The dashed lines in fig. \ref{S_states} correspond to the results of the separable model with a Gaussian radial function (eq. \ref{GaussianAnsatz}). The ground state almost coincides with the numerical evaluation, as expected. We have also checked numerically that the second isolated state is indeed a $2S$ state in the separable model (not shown).

\begin{figure}[!htbp]
\begin{center}\includegraphics[width=\columnwidth, keepaspectratio]{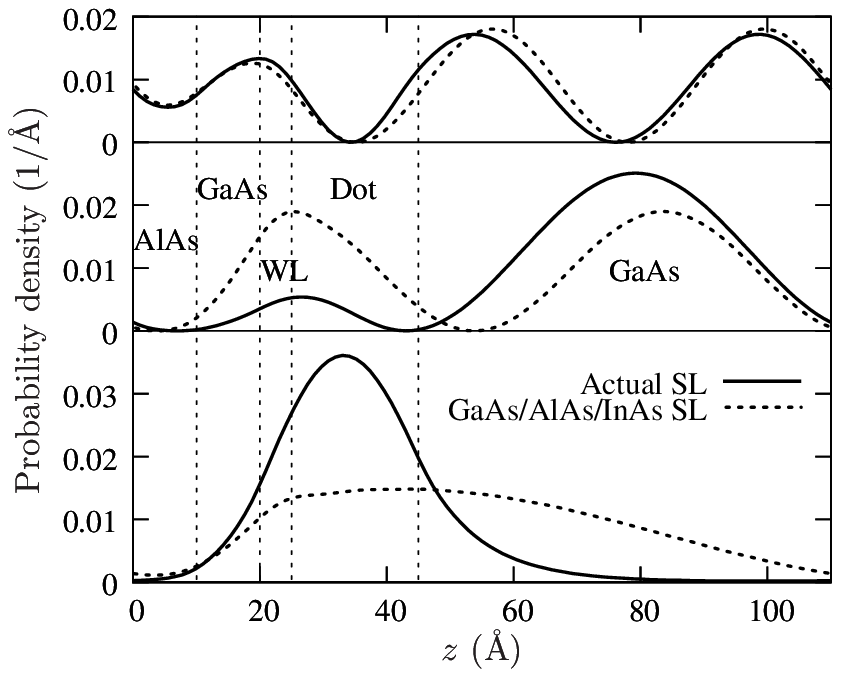}
\end{center}
\caption{Probability densities to find the electron at $z$ in the unit cell for the three lowest separable solutions with a Gaussian radial function and the three lowest states of a GaAs/AlAs/InAs SL at $B = 0$ and $k_z = 0$. From bottom to top: the ground state to the second excited state. The vertical dashed lines delimit the different regions of the structures in fig. \ref{dotmodel}}
\label{Probability}
\end{figure}

It is quite remarkable that several extrapolations of the fan chart to $B~=~0$, that is to say to the different edges of the SL of fig. \ref{dotmodel}, correspond to the energies of the excited separable states with no node in the layer plane \ref{GaussianAnsatz}. Let us compare the separable approximation of the actual SL in fig. \ref{dotmodel} to the GaAs/AlAs/InAs SL (i.e. with WL but without QDs). They differ by an extra kinetic energy $\hbar ^2 /(2m^* \sigma ^2)$ ($\approx$~40~meV) to be paid to localize laterally the electron in the dot on the one hand and an extra potential energy $\Delta V(z)$ which extends from the WL to the top of the dot and increases steadily from -0.4~eV to 0~eV \cite{Vasanelli01}. In a perturbative approach we would have to take the average of $\Delta V(z)$ over the probability densities associated with wave functions $c_{k_z=0}(z)$ or $c_{k_z=\pi/d}(z)$ of the GaAs/AlAs/InAs SL (see the dashed lines of fig. \ref{Probability} for $k_z~=~0$). For the ground state which is nodeless, the attractive contribution $\langle c_{k_z=0}|\Delta V| c_{k_z=0} \rangle$ is much greater than the lateral kinetic energy. As a result, the first edge of the GaAs/AlAs/InAs plunges from +19~meV down to -142~meV. Obviously, the first order is insufficient to quantitatively account for such a large effect of $\Delta V(z)$. In fact we note on fig. \ref{Probability} that the wave function of the separable model is considerably more concentrated on the dot along $z$ than the GaAs/AlAs/InAs ground state. Going up in energy the second edge (+151~meV) has an unperturbed $c_{k_z=0}(z)$ that has a node nearly at the center of the period (middle panel in fig. \ref{Probability}, dashed line). Hence, the $\Delta V$ attraction is considerably smaller than for the ground state, which leads to a much smaller shift between the two models. A similar situation holds for the excited states. Concomitantly, we see in fig. \ref{Probability} that the wave functions of the separable model become closer and closer to those of the GaAs/AlAs/InAs SL.

The appearance of low lying bound states and high energy resonances (dashed lines in fig. \ref{S_states}; see below) are an example of the deep attractive perturbation that the InAs QD brings to the energy spectrum of the GaAs/AlAs/InAs SL. Conversely, the continuum of the InAs QD is considerably restructured by the SL effect. Because of the small dot height, the resonances (or virtual bound states) for the $z$ motion occur at very high energy. Roughly, one can identify a QD to a quantum well with thickness $h$. Thus, we expect the resonance for the $z$ motion to occur at $\pi^2 \hbar^2 /(m^* h^2)$ from the InAs edge, i. e. in the eV range. Hence, one should expect a rather structureless continuum (resonances for the lateral motion will be discussed in the next section). Instead, the periodicity along $z$ gives rise to an energy spectrum that is a periodic function of $k_z$ and to wave functions that obey the Bloch theorem (eq. \ref{Bloch}). Moreover, the edge of the continuum of states is blue-shifted from -15~meV for an isolated QD (bottom of the thin WL quantum well) to about +16~meV in the periodic structure. The continuum states of the QDs and their WLs inserted into a GaAs/AlAs SL are therefore unique and cannot be simply considered as the result of the perturbation of one of the scheme (say the SL) by the other (the QD and its WL).

We show in fig. \ref{Probability} (full lines) the $z$ variations of the probability densities for the three lowest lying separable solutions:
\begin{equation}
P^{(n)} _{m=0}(z) = |y^{(n)} _{m=0}(z)|^2, \hspace{0.8cm} n = 1, 2, 3, \dots 
\end{equation}

These probability densities are extremely close to the actual numerical solutions but have the advantage of displaying unambiguous nodes while the latter is blurred in the exact solutions because of the slight non-separability. The departure from exact separability makes that the nodes of the actual problem follow curves $\psi_{k_z=0}(\rho,z)~=~0$. When $|\psi_{k_z}|^2$ is integrated over $\rho$ to produce the $P^{(n)} _{m=0}(z)$ of the actual sample, the integration over the radial variables changes a node line into a minimum along $z$. The closer this minimum is to zero, the more separable the problem is. $P^{(1)} _{m=0}$ is nodeless, $P^{(2)} _{m=0}$ has one node, $P^{(3)} _{m=0}$ has two nodes. While the ground solution is deeply bound to the dot, the two excited solutions leak significantly outside the dot.

If the separable model were exact, the Landau levels which belong to different zone center edges would cross at fields such that :
\begin{equation}
n \hbar \omega _c + E_l = n' \hbar \omega _c + E_{l'}
\end{equation}

where $E_l$ are the eigenenergies for the $z$ motion.

\begin{figure}[!htbp]
\begin{center}\includegraphics[width=\columnwidth, keepaspectratio]{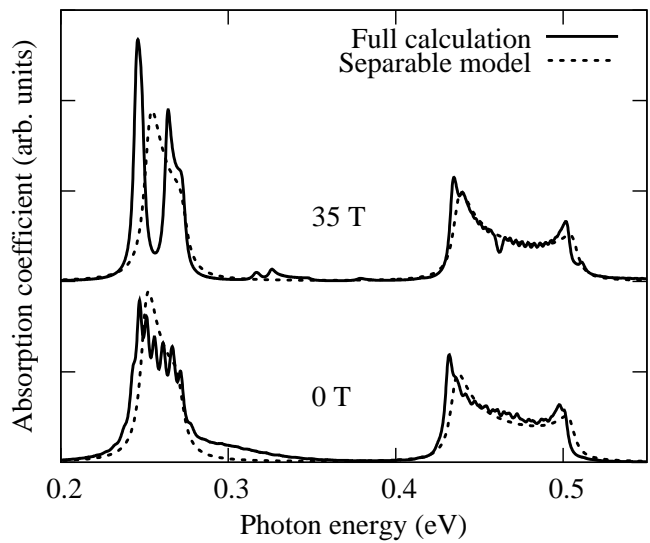}
\end{center}
\caption{Absorption coefficients versus photon energy at $B~=~0$ and $B~=~35~\mathrm{T}$ calculated by two models for $\vec{\mathcal{E}}$ parallel to $\vec{z}$}.
\label{PolaZComp}
\end{figure}

We see in fig. \ref{S_states} that these crossings are actually replaced by anti-crossings. Their magnitude gives a measure of the inaccuracy of the separable ansatz. Since the QDs are flat, a great deal of the continuum states can be analyzed in terms of separable states. The states predicted by the separable model play an important role in optical absorption as shown in fig. \ref{PolaZComp}. To obtain these results, a summation over 60 $k_z$ values of the $k_z$-conserving transitions between the initial and final states has been performed. The coincidence of the two calculated absorption coefficients for the $z$ polarization, using either the full 3D calculation or the separable model at $B = 0$ or $B = 35~\mathrm{T}$, evidences that the ground state couples preferentially with the continuum states which have nearly the same in-plane extension. Hence, the remainder of the continuum states contribute very little to the optical absorption. Note that the existence of the two peaks in the full calculation and only one in the separable model for the first absorption feature at 35~T is due to the non-separability of the total wave function, in correspondence with the anti-crossing illustrated by large circles in fig. \ref{S_states}. Note finally that the absorption spectrum reflects the presence of minibands: their widths are equal to $E_2(k_z=0) - E_2(k_z=\pi/d) \approx 40~\mathrm{meV}$ and $E_3(k_z=\pi/d) - E_3(k_z=0) \approx 80~\mathrm{meV}$ for the first and second features respectively (the width of the ground miniband is negligible) while their double-peak profiles reflect the singular density of states at the edges of a SL miniband.

\begin{figure}[!htbp]
\begin{center}\includegraphics[width=\columnwidth, keepaspectratio]{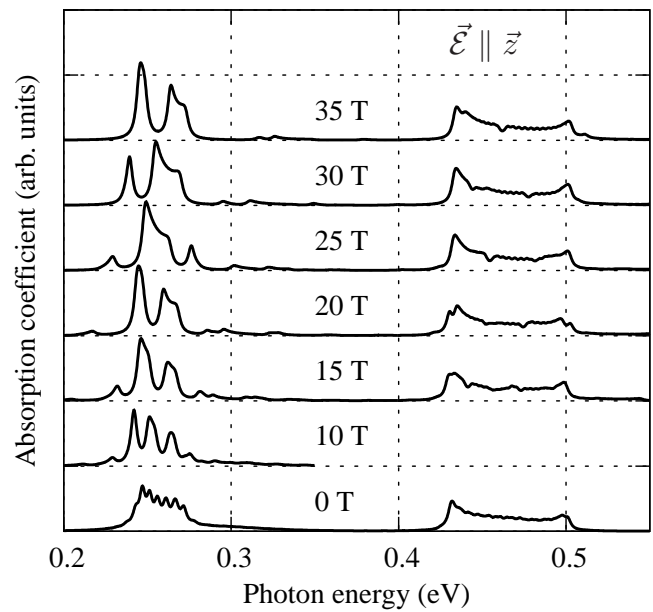}
\end{center}
\caption{Absorption coefficients versus photon energy from $B = 0$ to $B =35~\mathrm{T}$ for $\vec{\mathcal{E}}$ parallel to $\vec{z}$.}
\label{PolaZ}
\end{figure}

We show in fig. \ref{PolaZ} the calculated absorption coefficients from $B = 0$ to $B = 35~\mathrm{T}$ for light polarized parallel to $z$. The spectrum at $B = 10~\mathrm{T}$ is cut off at 0.35~eV due to computational limits. We immediately see that the spectrum is almost $B$ independent, in spite of a quasi-fan chart aspect of the continuum levels (fig. \ref{S_states}). The absorption takes place between the strongly localized initial state and the continuum states which `look' similar to the ground state in the layer plane, i. e. the separable excited states (which form a 1D continuum with frozen in-plane motion and periodic $z$ motion; note that the separable model describes excellently the second profile around 0.45~eV while the marked oscillations in the first profile are due to non-separability effects). The transitions towards all the other states of the fan charts in fig.~\ref{S_states} are very faint because of the mismatched in-plane variations between the initial and final states. The insensitivity to an external field of the absorption is a direct consequence of the quasi-separability of the vertical and in-plane motions. In this respect the QDs behave much the same as narrow one dimensional square wells. In these systems, it is known that for $\vec{\mathcal{E}}$ and $\vec{B} \parallel \vec{z}$ the intersubband absorption is also roughly $B$ independent (if it were not for band non parabolicity and/or effective mass mismatch there would be an exact decoupling between the vertical and in-plane motions and thus an exact parallelism between the $n^{th}$ Landau level of all the $z$ dependent subbands).

\begin{figure}[!htbp]
\begin{center}\includegraphics[width=\columnwidth, keepaspectratio]{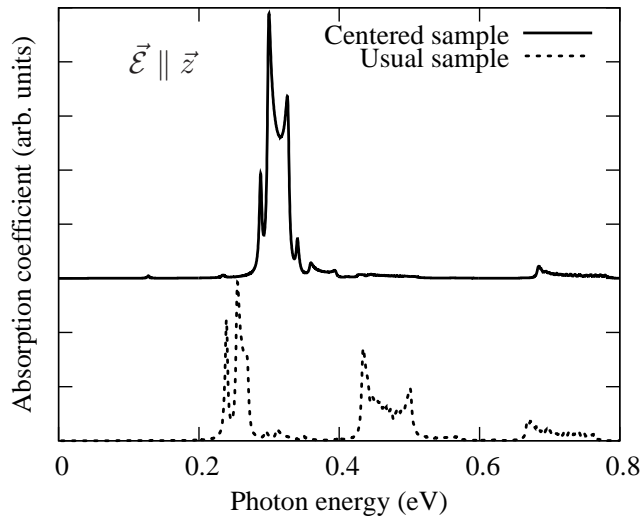}
\end{center}
\caption{Absorption coefficients versus photon energy at $B~=~30~\mathrm{T}$ for $\vec{\mathcal{E}}$ parallel to $\vec{z}$. Usual sample and a centered sample (see text). The full line has been rigidly up-shifted for clarity.}
\label{MiddleDot}
\end{figure}

Fig. \ref{MiddleDot} compares the absorption coefficients for two samples: the usual sample has been described above while the centered sample has the same material parameters except that the QD has been placed at the center of the SL period. By comparing to fig. \ref{PolaZ}, we see that the second absorption peak has disappeared while the first one has been reinforced. By placing the QD, which is a flat object, at the center of the SL cell, we expect to restore a quasi-mirror symmetry with respect to the center of the period, thereby making transitions from the ground state (which is even in $z$) towards even excited states almost forbidden in the $z$ polarization.

To conclude this section, let us emphasize that the absorption peaks above are strong and observable in experiments. At a given magnetic field $B$ and wave vector $k_z$, the sum of oscillator strength of the transitions from the ground state to the calculated continuum states amounts to $\sim$~0.8. This is close to the saturation value ($=1$) stated by Thomas-Kuhn-Reich \cite{Thomas} sum rule. With a structure which has 20 layers and a lateral QD density of $5\times 10^{10}\mathrm{cm}^{-2}$, these peaks were observed either in photocurrent \cite{Rebohle02} or in absorption \cite{Sauvage97} experiments.

\subsection{$P$ states and excitation with in-plane polarization}

\begin{figure}[!htbp]
\begin{center}\includegraphics[width=\columnwidth, keepaspectratio]{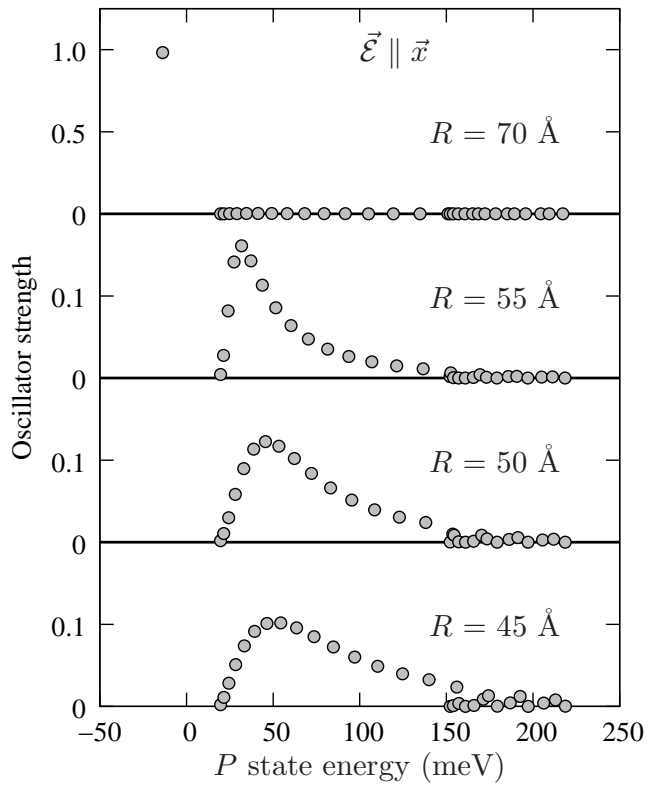}
\end{center}
\caption{Oscillator strength of transitions from the ground $S$ state to the first 30 $P$ states versus energy of $P$ states at $B=0$ and $k_z = 0$ for several values of base radius $R$ and $\vec{\mathcal{E}} \parallel \vec{x}$. The ordinate scale in the case $R~=~70~\mathrm{\AA}$ is 5 times bigger than the others. The ground energies for each value of $R$ are: $E_S(R=70~\mathrm{\AA})~=~-105~\mathrm{meV}$, $E_S(R=55~\mathrm{\AA})~=~-64~\mathrm{meV}$, $E_S(R=50~\mathrm{\AA})~=~-45~\mathrm{meV}$, $E_S(R=45~\mathrm{\AA})~=~-25~\mathrm{meV}$}
\label{OscillatorForce}
\end{figure}

\begin{figure}[!htbp]
\begin{center}
\includegraphics[width=\columnwidth, keepaspectratio]{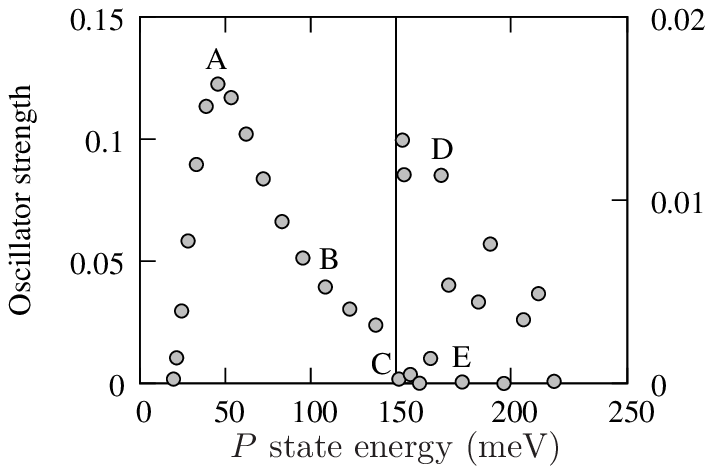}\\
\includegraphics[width=\columnwidth, keepaspectratio]{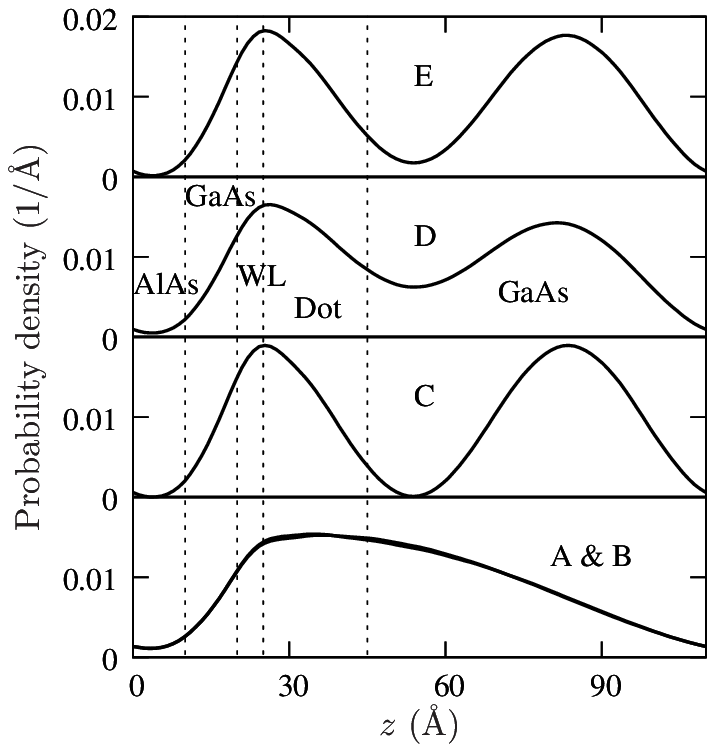}
\end{center}
\caption{Upper panel: Oscillator strength of transitions from the ground $S$ state to the first 30 $P$ states versus energy of $P$ states at $B=0$ and $k_z = 0$ for $R=50~\mathrm{\AA}$ and $\vec{\mathcal{E}} \parallel x$. Two scales of vertical axis are used below and above 150~meV. The apparent scattering of the numerical data above 150~meV is due to the contribution of two different optical channels (see text). Lower panel: Probability density to find the electron at $z$ in the unit cell of some selected states marked (from A to E) in the upper panel.}
\label{VirtualWave}
\end{figure}

When the QDs support bound states with $P\pm$ symmetry, the bound-to-bound $S \to P\pm$ absorption coefficients are very strong for $\vec{\mathcal{E}} \perp \vec{z}$. These two transitions exhaust almost all the oscillator strength and saturates the Thomas-Kuhn-Reich sum rule. Hence only the $z$ polarization gives rise to a strong bound-to-continuum transition (as discussed in section \ref{Section1}). A possible means to overcome the weakness of the absorption with in-plane polarization light is to push the $P$ levels in the continuum where they might form resonant states. Under such circumstances the $S$-to-continuum absorption will also be strong in the in-plane polarizations. This can be achieved by decreasing the QD base radius $R$ until the lowest $P$ state energy exceeds the onset of the continuum. For the ground $P$ level, this happens at about 5.8~nm. Since the QD has sharp boundaries, it is likely that the bound state will survive in the continuum in the form of a virtual bound state \cite{Lelong00, Bastard91}. Such a feature would be beneficial to the photoconductive properties of the device since, ideally speaking, a virtual bound state would have a large oscillator strength distributed over an energy interval broader than that of a true bound state and be at the same time able to participate in the electrical conduction. To track the evolution of the lowest lying continuum $P$ states of the QDs, we have calculated for several values of $R$ the integrated probability to find the electron in the QD (not shown) and the oscillator strength of the $S$-$P$ transitions when light is polarized along the $x$ axis:
\begin{equation}
  OS_{S \rightarrow P_x} = \frac{2m^*}{\hbar ^2} (\epsilon_{P_x} - \epsilon_{S}) |\langle P_x| x |S\rangle |^{2}
\end{equation}

Fig. \ref{OscillatorForce} shows the oscillator strength versus the energy of the first 30 final $P$ states for several $R$ values at $B=0$ and $k_z = 0$. For $R = 70~\mathrm{\AA}$, one $P$ state is bound. Thus, there is a very sharp peak at~-14~meV. It is followed by a smooth and very weak ($OS < 1.5~\%$) continuum absorption in the positive photon energy region. For smaller dots, there is no bound $P$ state. Consequently, the oscillator strength becomes quite strong for final states in the continuum spectrum. A peak develops at an increasingly larger energy when $R$ decreases. Its width increases with decreasing dot size. We note that the sum of the oscillator strength inside each broad peak amounts to $\sim$ 0.9, a value close to the bound-to-bound transition. This suggests that the oscillator strength has been redistributed over the peak. At a $P$ state energy of about 150~meV, independently of $R$, one notes the existence of two kinds of states. A first kind of state appears as the continuation of the absorption peak while the other kind of state are basically optically inactive. We show on fig. \ref{VirtualWave} the evolution of the $z$ dependence for the probability density:
\begin{equation}
P_r (z) = \int _0 ^\infty 2 \pi \rho \mathrm{d} \rho |\psi _{P_x} (\rho, z)|^2
\end{equation}

with increasing states $n$. While the states that belong to the absorption peak have no node along $z$, the states with a vanishing oscillator strength show one node along $z$. Hence, the apparent cut-off at 150~meV corresponds to the onset of a new (but very small) absorption channel, this time associated with a change of the quantum number of the $z$ motion. We note that the first two eigenenergies of a SL with GaAs, AlAs, InAs WL but without InAs QDs are 19~meV and 151~meV at $B=0$ and $k_z = 0$ (see discussions for $S$ states in section \ref{Section1}), that is to say the two energies of the absorption onsets shown in fig. \ref{OscillatorForce}. This shows once again that the flat aspect of the QDs leads to a quasi-separability of the carrier motions along and perpendicular to the growth axis.

\section{Conclusions}

In conclusion, we have calculated the electronic continuum states of InAs/GaAs QDs embedded into a GaAs/AlAs SL in the presence of a quantizing magnetic field applied parallel to the growth axis. We find that the far infrared absorption is essentially independent of the vertical magnetic field despite the presence of numerous quasi-Landau states. We have proved that the flatness of the QDs plays a major part in restructuring the QD's continuum. Moreover, we have shown that small dots (with a single bound state of $S$ symmetry) may display a strong in-plane absorption when a virtual bound state with $P$ symmetry is not too far from the continuum edge. Finally, we have shown that most of the optical properties of QDs embedded in a SL can be analyzed within a decoupled model. This results from the flatness of the QD shape and from the fact that the dipole matrix elements are sensitive only to the local (i.e. in the QD region) features of the final state wave functions, because of the strong localization of the initial QD state.

Let us finally add a few comments regarding QDs with several electrons. It is known that QD can contain several electrons. When several electrons exist in the dots, the Pauli principle may block bound-to-bound transitions if the shell of final states is filled. In addition, the many electron states in the continuum comprise not only states which can correspond to the delocalized states for all electrons but also mixed states where some electrons are bound to the dot while the others are extended. For instance, in the two electron case, the initial state (at low temperature) is a spin singlet with the two electrons in the ground state. The final states can be bound for both electrons or mixed with one electron bound and one in the continuum. As a result of these mixed states, the $(S-S)-(S-P)$ bound-to-bound (discrete) transitions will be accompanied by $(S-S)-(S-$continuum) continuum transitions. The Coulomb interaction will be more important in the bound-to-bound transitions than in the mixed transitions. Hence, the $(S-S)-(S-P)$ discrete lines can be very close in energy to the $(S-S)-(S-$continuum) absorption threshold. The description of the absorption coefficient of QDs containing more than one electron is clearly beyond the scope of this paper.

\begin{acknowledgments}
The LPA-ENS is `laboratoire associ\'e au CNRS UMR 8551 et aux Universit\'es Pierre et Marie Curie (Paris 6) et Denis Diderot (Paris 7)'. We are grateful to F.F.~Schrey, G.~Strasser and K.~Unterrainer for helpful discussions. One of us (G. B) acknowledges the Wolfgang Pauli Foundation for partial support.
\end{acknowledgments}


\begin{thebibliography}{10}
\bibitem{Thomas}	The Thomas-Kuhn-Reich sum rule states that the sum of oscillator strength for transitions from an eigenstates to all the other eigenstates in a given polarization is equal to 1.
\end{thebibliography}
\end{document}